\documentclass[aps, prd, 10pt, twocolumn, superscriptaddress, noshowpacs, preprintnumbers, longbibliography, groupedaddress, footinbib, bibnotes]{revtex4-1}

\usepackage{amsmath}
\usepackage{amsfonts}
\usepackage{amssymb}
\usepackage{bbold}
\usepackage{epsfig}
\usepackage{graphicx}
\usepackage{bm}
\usepackage{array}
\usepackage{hyperref}
\usepackage{listings}
\usepackage{color}
\usepackage{float}
\usepackage[normalem]{ulem}

\newcommand{\Closs}{\mathcal{C}_{\textrm{loss}}}
\newcommand{\Cgain}{\mathcal{C}_{\textrm{gain}}}
\newcommand{\Clossbar}{\bar{\mathcal{C}}_{\textrm{loss}}}
\newcommand{\Cgainbar}{\bar{\mathcal{C}}_{\textrm{gain}}}

\begin{document}

\title{The three  flavor revolution in fast pairwise neutrino conversion}

\author{Shashank Shalgar}
\email{shashank.shalgar@nbi.ku.dk}
\thanks{ORCID: \href{http://orcid.org/0000-0002-2937-6525}{0000-0002-2937-6525}}
\affiliation{Niels Bohr International Academy \& DARK, Niels Bohr Institute,\\University of Copenhagen, 2100 Copenhagen, Denmark}

\author{Irene Tamborra}
\email{tamborra@nbi.ku.dk}
\thanks{ORCID: \href{http://orcid.org/0000-0001-7449-104X}{0000-0001-7449-104X}}
\affiliation{Niels Bohr International Academy \& DARK, Niels Bohr Institute,\\University of Copenhagen, 2100 Copenhagen, Denmark}

\date{\today}

\begin{abstract}
The modeling of  fast flavor evolution of neutrinos in dense environments has been traditionally carried out by relying on a two flavor approximation for simplicity. In addition, vacuum mixing has been deemed  negligible. For the first time, we highlight that 
the fast flavor evolution in three flavors is intrinsically different from the one obtained in the two flavor approximation.  This is due to the exponential growth of  flavor mixing in the $e$--$\mu$ and $e$--$\tau$ sectors generated by the vacuum term in the Hamiltonian. As a result, substantially larger flavor mixing is found in three flavors.  Our findings highlight that the two flavor approximation is not justified for fast pairwise conversion, even if the angular distributions of non-electron type neutrinos are initially identical.  
\end{abstract}

\maketitle

\section{Introduction}
\label{sec:intro}

Flavor mixing is one of the least understood phenomena  affecting the physics of neutrino-dense astrophysical sources, such as core-collapse supernovae and compact binary mergers. The  flavor evolution can be significantly modified by the presence of a background medium, as first proposed by Mikheev, Smirnov, and Wolfenstein~\cite{Mikheev:1986if,1985YaFiz..42.1441M,1978PhRvD..17.2369W}. In addition, in neutrino-dense media, the forward scattering of neutrinos off their surrounding neutrino ensemble can crucially modify the flavor evolution in a non-linear fashion~\cite{Pantaleone:1992eq, Sigl:1992fn}. 
The flavor evolution induced by neutrino-neutrino interactions  is correlated and often referred to as {\it collective} in nature. It is such that neutrinos of different energies change their flavor coherently.

More recently, it has been pointed out that the pairwise forward scattering of neutrinos of the form $\nu_e \bar\nu_e \leftrightarrow \nu_x \bar\nu_x$ (with $x=\mu, \tau$) can trigger collective flavor conversions, if the neutrino density is large enough~\cite{Sawyer:2005jk, Sawyer:2008zs, Sawyer:2015dsa,Chakraborty:2016yeg,Chakraborty:2016lct,Tamborra:2020cul}. A necessary, although not sufficient, condition leading to  the development of flavor mixing is the occurrence of  crossings in the electron neutrino lepton number (ELN)~\cite{Izaguirre:2016gsx}.  The development of fast pairwise conversions has been deemed to be not linked to the vacuum frequency of neutrinos; hence, pairwise conversion of neutrinos could occur even if the atmospheric neutrino mass-squared difference $\Delta m^2 = 0$ and should be independent on the neutrino energy $E$. However, it is shown in Ref.~\cite{Shalgar:2020xns} that the vacuum frequency $\Omega=\Delta m^2/2E \neq 0$ may strongly affect flavor mixing in the non-linear regime. Understanding the implications of fast pairwise conversion on the final flavor outcome is of relevance because fast conversions could have implications on the supernova explosion mechanism and on the synthesis of the heavy elements~\cite{Xiong:2020ntn, Wu:2017drk, George:2020veu,Tamborra:2017ubu,Shalgar:2019kzy,Abbar:2018shq, Azari:2019jvr,DelfanAzari:2019tez,Glas:2019ijo,Abbar:2019zoq,Nagakura:2019sig,Morinaga:2019wsv,Capozzi:2020syn,Abbar:2020qpi}.

In order to grasp the behavior of neutrino collective oscillations, in analytical treatments and numerical simulations, a two flavor $(\nu_e,\nu_x)$ approximation has been classically adopted, as also considered above. This assumption was justified by the fact that significant flavor evolution of neutrinos is possible, if the neutrino self-interaction potential $\mu = \sqrt{2} G_{\textrm{F}} n_{\nu}$ (with $G_\textrm{F}$ being the Fermi constant and $n_\nu$ the neutrino number density) is much larger than the typical vacuum frequency $\Omega=\Delta m^2/2E$; the latter, in turn, is larger than the vacuum frequency due to the solar mass difference $\omega=\delta m^2/2E$ [note that two mass-squared differences differ by a factor $\mathcal{O}(30)$]. In addition, due to the typical energies of these astrophysical environments, the  emission properties of $\nu_\mu$ and $\nu_\tau$  were assumed to be identical~\cite{Duan:2005cp, Duan:2006an, Duan:2006jv, Duan:2007bt, Duan:2010bg}. 
As for slow neutrino self-interaction (exhibited by systems that are stable in the limit $\omega, \Omega\rightarrow 0$), the effects of  $\Omega$ and $\omega$ on the collective flavor evolution could be factorized and the overall two flavor approximation has been deemed to be  predictive of the final flavor outcome~\cite{Fogli:2008fj,Dasgupta:2010cd,Dasgupta:2010ae,Dasgupta:2007ws,Dasgupta:2008cd,Friedland:2010sc}.   In the case of fast flavor conversion, both vacuum frequencies have been  neglected under the assumption that flavor mixing would be completely driven by the (anti)neutrino number density.

The  $\nu_{\mu,\tau}=\bar\nu_{\mu,\tau}$ assumption has been  challenged by the most recent supernova simulations allowing for muon production~\cite{Bollig:2017lki}, and  crossings have been found in the non-electron lepton number (nELN) distributions of neutrinos~\cite{Capozzi:2020syn}.  
These developments have drawn attention  on possible three flavor effects linked to fast pairwise conversion~\cite{Airen:2018nvp,Chakraborty:2019wxe}. In particular, Ref.~\cite{Capozzi:2020kge} has recently pointed out that the existence of large  nELN crossings  can vastly affect the flavor outcome.

In this work, we build on the findings of Ref.~\cite{Shalgar:2020xns} and expand on the ones of Refs.~\cite{Airen:2018nvp,Chakraborty:2019wxe,Capozzi:2020kge}.  For the first time, highlight the surprisingly strong dependence of the final flavor outcome on $\Omega$ and $\omega$   in the non-linear regime, even in the absence of zero nELN crossings considered in Ref.~\cite{Capozzi:2020kge}; this may seem, at first, counter-intuitive, since $\omega \ll \Omega \ll \mu$. Our work highlights the limits of the  usually adopted two flavor approximation and the strong dependence  on vacuum flavor mixing:  the main qualitative difference between the three and two flavor cases arises because the difference between  $\nu_{\mu}$ and $\nu_{\tau}$ in the three flavor case grows exponentially, even if they are assumed to be  identical  or isotropic initially. This pins down the essence of three flavor effects and was not appreciated in Refs.~\cite{Airen:2018nvp,Chakraborty:2019wxe,Capozzi:2020kge}. In addition,  the non-trivial angular distributions for the  non-electron type neutrinos effectively change the  lepton number distribution and the onset of flavor mixing.

This paper is organized as follows. In Sec.~\ref{setup}, we introduce the equations of motion and formulate the problem setup. In Sec.~\ref{sec:results}, we compare the three flavor outcome to the two flavor scenario for the case where the non-electron neutrinos are only created through flavor mixing and for the scenario where crossings exist in the initial electron and muon neutrino lepton number distributions; we also explore the dependence of the flavor conversion physics on the neutrino mixing parameters. In Sec.~\ref{sec:collisions}, the effect of collisions on the flavor mixing phenomenology in three flavors is discussed. 
An outlook on our findings  and conclusions are reported  in Sec.~\ref{sec:conclusions}. The numerical convergence of the results presented in this work is discussed in Appendix~\ref{sec:convergenge}, together with the employed numerical techniques. The effect of the matter background on flavor mixing is discussed in Appendix~\ref{sec:matter}.

\section{Equations of motion and initial system configuration}
\label{setup}
The equations of motion are formally identical  in  two and three flavor systems--apart from the degrees of freedom. For each momentum mode $\vec{p}$ and in the two flavor approximation, the flavor state can be represented by a $2 \times 2$ density matrix for neutrinos ($\rho$) and similarly for antineutrinos ($\bar\rho$). In the three flavor case, the density matrices are promoted to $3 \times 3$ matrices which evolve in accordance to the following equations of motion in the mean-field approximation,
\begin{eqnarray}
\label{eq:eom1}
i\frac{d}{dt} \rho(\vec{p}) = [H(\vec{p}),\rho(\vec{p})] \ \mathrm{and}\ 
i\frac{d}{dt} \bar{\rho}(\vec{p}) = [\bar{H}(\vec{p}),\bar{\rho}(\vec{p})]\ .
\end{eqnarray}
We only consider an homogeneous neutrino gas, hence the total derivative with respect to time can be replaced by a partial derivative.
The Hamiltonian receives  contributions from the vacuum, matter, and the self-interaction terms: 
\begin{equation}
H = H_{\textrm{vac}} + H_{\textrm{mat}} + H_{\nu\nu} \ ,
\end{equation}
with
\begin{eqnarray}
H_{\textrm{vac}} &=& \frac{1}{2E} U M^{2} U^{\dagger}\ ,\\
H_{\textrm{mat}} &=& 
\textrm{diag}(\lambda, 0, 0)\ ,\\
H_{\nu\nu}\ &=& \mu \int d^{3} \vec{p^{\prime}} [\rho(\vec{p^{\prime}})-\bar{\rho}(\vec{p^{\prime}})] \left(1-\frac{\vec{p}\cdot\vec{p^{\prime}}}{|\vec{p}||\vec{p^{\prime}}|}\right)\ .
\end{eqnarray}
The matrix $M$ represents the diagonal mass matrix, while $U(\theta_{12}, \theta_{13}, \theta_{23}, \delta)$ is the PMNS matrix in the standard parametrization~\cite{Zyla:2020zbs}. The matter potential  is $\lambda=\sqrt{2} G_{\textrm{F}} n_{e}$, where  $n_{e}$ is the number density of electrons, and  $\mu$ is the effective strength of the neutrino self-interaction, which is proportional to the neutrino number density. 
For antineutrinos, the Hamiltonian $\bar{H}$ has the same form as $H$, with $H_{\textrm{vac}}$ being replaced by $-H_{\textrm{vac}}^{\star}$. 

In the numerical simulations, we assume the mixing parameters entering $H_{\textrm{vac}}$ reported in Table~\ref{Table1}. In the following, we intend to compare the three flavor outcome with the two flavor one. To this purpose, we note that the three flavor system reduces to the two flavor approximation for  $\theta_{12}=0$, $\theta_{23}=0$, and $\omega=0$.
\setlength{\extrarowheight}{3pt}
\begin{table}
\caption{Default values of the vacuum mixing parameters adopted in our numerical simulations as from Ref.~\cite{Zyla:2020zbs} for normal mass ordering.}
\label{Table1}
\begin{tabular}{|l|l|}
\hline
Parameter & value\\
\hline
\hline
$\delta m^{2}$ & $7.53 \times 10^{-5}$ eV$^{2}$ \\
\hline
$\Delta m^{2}$ & $2.45 \times 10^{-3}$ eV$^{2}$ \\
\hline
$\theta_{12}$ & $33.65^{\circ}$ \\
\hline
$\theta_{13}$ & $8.49^{\circ}$ \\
\hline
$\theta_{23}$ & $47.58^{\circ}$ \\
\hline
$\delta$ & $0.0$ \\
\hline
\end{tabular}
\end{table}

In the two flavor approximation, it is common to ignore the matter effect and  instead use a small mixing angle to take into account  matter suppression effects~\cite{EstebanPretel:2008ni}. This, however, has not been tested in the three flavor case in the context of fast conversions;  hence, our numerical simulations include the matter potential and assume $\mu = \lambda = 10^5$~km$^{-1}$, unless otherwise stated.

Apart from homogeneity, we also consider a system that is axially symmetric and, for simplicity, assume that the neutrino gas contains (anti)neutrinos with $E=50$~MeV to  mimic the neutrino average energies just before decoupling. The neutrino momentum can thus be replaced by the cosine of the polar angle, $\cos\theta$, and both the density matrices and the Hamiltonian are independent of the spatial coordinates.

\begin{figure}[b]
\includegraphics[width=0.49\textwidth]{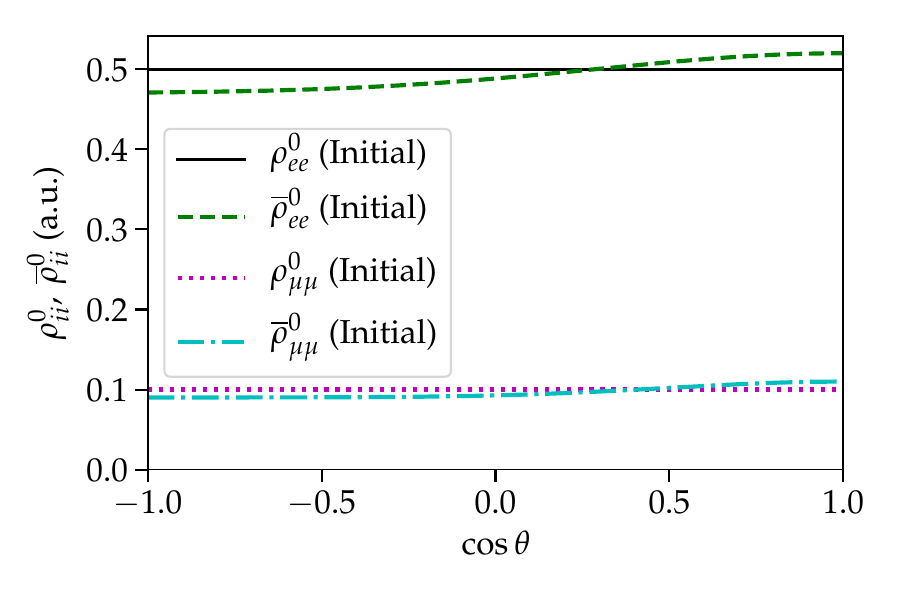}
\caption{Initial angular distributions of electron neutrinos and antineutrinos. We assume an isotropic distribution for $\nu_e$ (black solid line) and a forward peaked distribution for $\bar\nu_e$ (dashed green line). 
For Case I, we set the distributions of  non-electron type neutrinos to zero. For Case II, we use the angular distributions shown using the dotted and dash-dotted lines for the muon flavors with $a=10^{-2}$, while the distributions of the tau flavors remain zero.} 
\label{Fig1}
\end{figure}

The fast flavor evolution depends on the angular distribution of (anti)neutrinos. At $t=0$
\begin{subequations}
\begin{eqnarray}
\label{cases1}
 \rho_{ee}^0(\cos\theta) &=& 0.5  \\ 
  \bar{\rho}_{ee}^0(\cos\theta) &=& 0.47 + 0.05 \exp\left[-(\cos\theta-1)^{2}\right] \ ,
\label{cases2}
\end{eqnarray}
\end{subequations}
as displayed in Fig.~\ref{Fig1}. 

For the non-electron flavors,  we consider two scenarios: one  with  the angular distributions  for non-electron (anti)neutrinos initially equal to zero (Case I) and one with the non-electron neutrino distributions different from zero (Case II). Hence, for Case I:
\begin{subequations}
\begin{eqnarray}
\label{eq:zero}
\rho_{\mu\mu}^0(\cos\theta) &=&  \bar\rho_{\mu\mu}^0(\cos\theta) = 0\\
 \rho_{\tau\tau}^0(\cos\theta) &=&  \bar\rho_{\tau\tau}^0(\cos\theta) = 0\ .
 \label{eq:zero1}
\end{eqnarray}
\end{subequations}
For Case II:
\begin{subequations}
\begin{eqnarray}
\label{mucr1}
 \rho_{\mu\mu}^0(\cos\theta) &=& 0.1 \\ 
 \bar{\rho}_{\mu\mu}^0(\cos\theta) &=&  (0.1 - a) +  2 a \exp\left[-2 (\cos\theta-1)^{2}\right]\ , 
\label{mucr2}
\end{eqnarray}
\end{subequations}
where the parameter $a$ determines the magnitude of the crossing in the muon type neutrinos and we  assume $a=0, 10^{-3}$, and $10^{-2}$; see Fig.~\ref{Fig1}. In Case II, we assume  $\rho_{\tau\tau}^0(\cos\theta)=\bar{\rho}_{\tau\tau}^0(\cos\theta) = 0$. For the interested reader, details on the numerical solution of the neutrino equations of motion and their convergence are provided in Appendix~\ref{sec:convergenge}.

\section{Flavor mixing in two versus three flavors}
\label{sec:results}
In this section, we explore the flavor mixing in the two and three flavor scenarios for Cases I and II. We also discuss the effect of the vacuum mixing parameters on the flavor conversion physics.
\subsection{Case I: Non-electron flavors are produced through flavor mixing}
We  explore the difference between the two and three flavor cases  by assuming  Eqs.~\ref{cases1}, \ref{cases2}, \ref{eq:zero}, and \ref{eq:zero1} as initial conditions.  
In order to allow for a fair comparison between the two and three flavor cases, we introduce the angle-averaged ratio between the initial and final content of $\nu_e$: 
\begin{eqnarray}
\langle \tilde{P}_{ee} \rangle = \frac{\int \rho_{ee}(\cos\theta,t) d\cos\theta}{\int \rho_{ee}^0(\cos\theta) d\cos\theta}\ ;
\end{eqnarray}
this ratio coincides with the angle-averaged $\nu_e$ survival probability, if $\rho_{\mu\mu}^0(\cos\theta)=\rho_{\tau\tau}^0(\cos\theta)$.

\begin{figure*}
\includegraphics[width=0.49\textwidth]{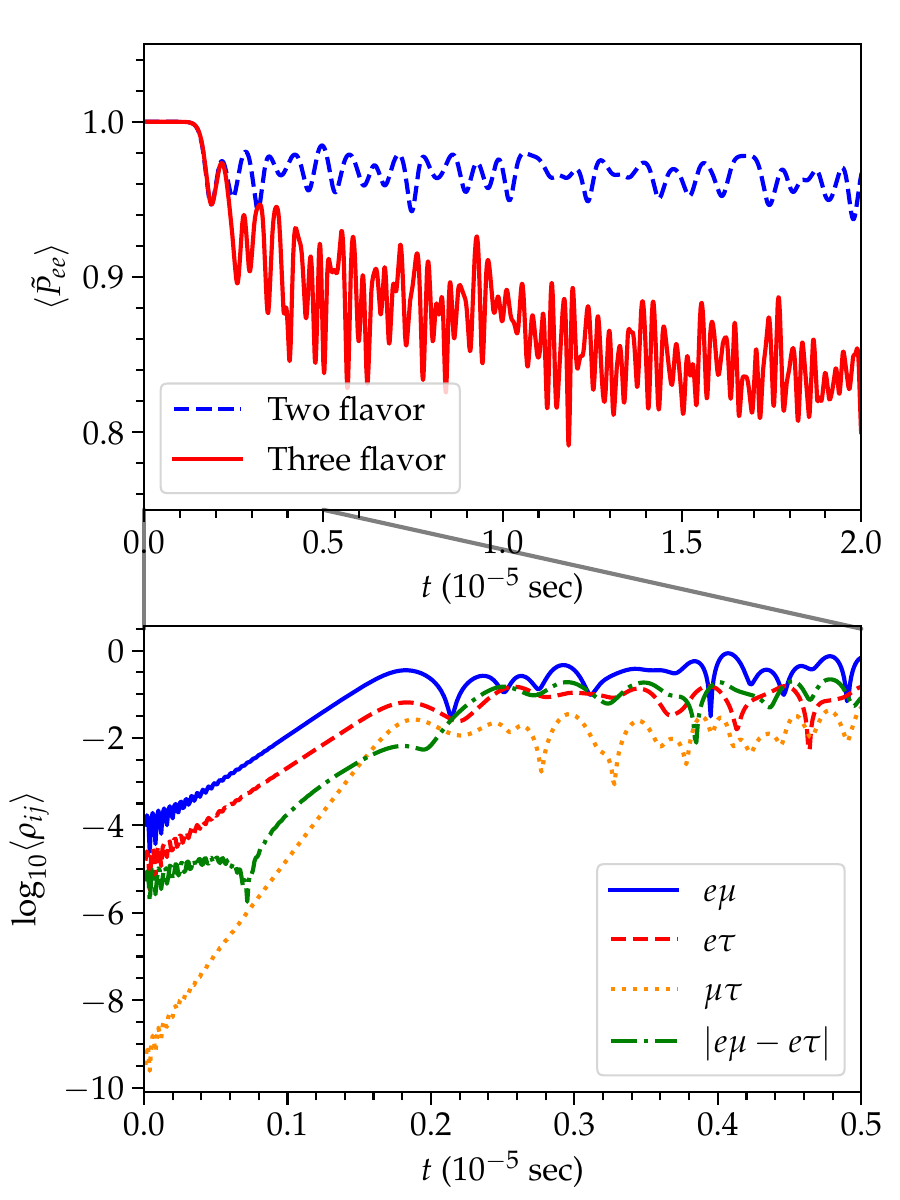}
\includegraphics[width=0.49\textwidth]{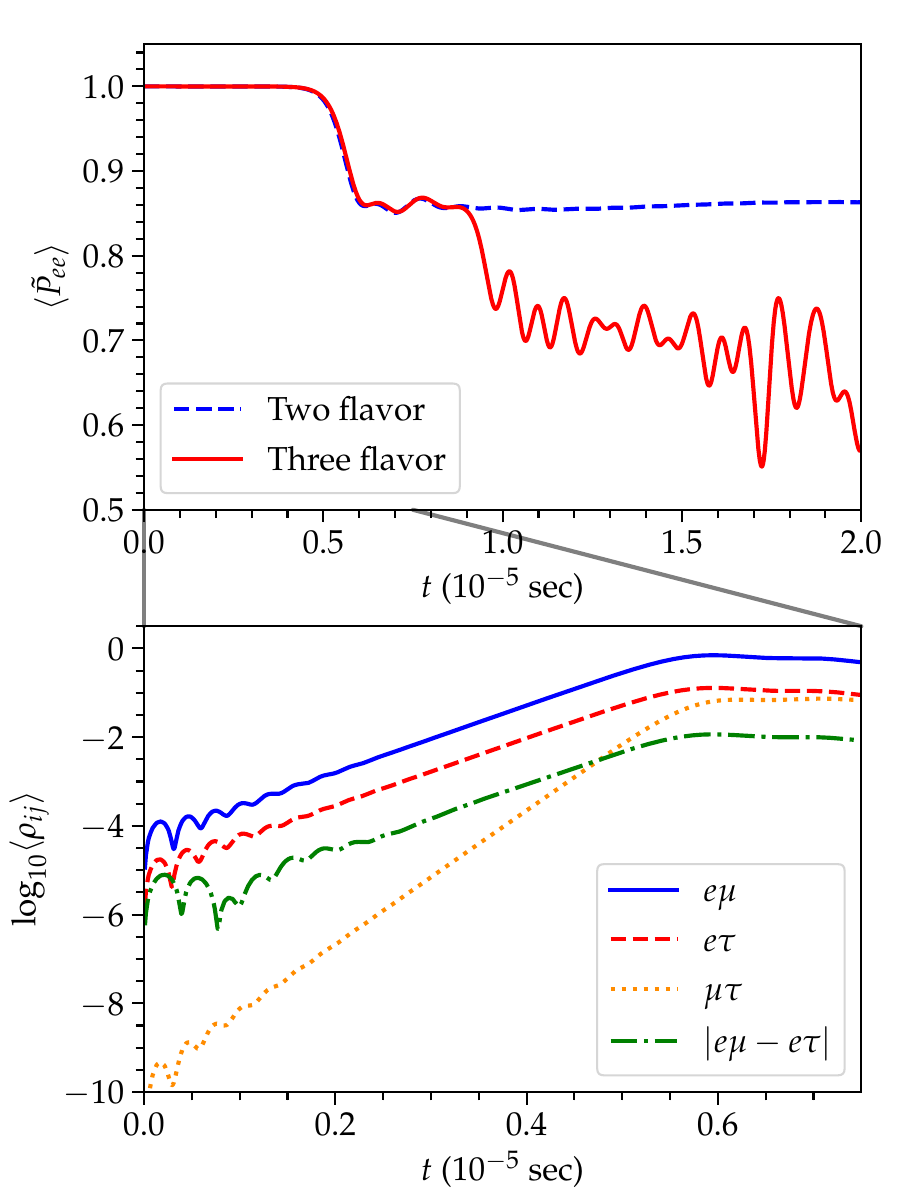}
\caption{{\it Left:} Angle averaged ratio between the initial and final content of $\nu_e$  for Case I (see Eqs.~\ref{cases1} and \ref{cases2})  as a function of time for $\mu = \lambda = 10^{5}$ km$^{1}$ (blue dashed line for the two flavor approximation and  red solid line for the  three flavor case). In the inset on the bottom, the temporal evolution of the absolute values of the angle averaged off-diagonal elements of $3 \times 3$ density matrix is shown.  {\it Right:} Same as the left panels,  but for  
$\mu=10^{4}$ km$^{-1}$. In both cases, the growth of $|\rho_{e\mu} - \rho_{e\tau}|$ in three flavors is responsible for the difference with respect to the two flavor case. The case with the  smaller value of $\mu$ displays a growth rate for the off-diagonal elements that is smaller as expected from the linear stability analysis; however, the total conversion rate is larger (note the different ranges on the $y$-axis of the top panels). } 
\label{fig:fig2}
\end{figure*}

Figure~\ref{fig:fig2}  shows the temporal evolution of the angle averaged ratio between the initial and final content of $\nu_e$  in the two (dashed blue line) and three (solid red line) flavor cases.  
The left panels of Fig.~\ref{fig:fig2} are obtained for our default values of $\lambda$ and $\mu$ ($\lambda = \mu=10^{5}$ km$^{-1}$), while the right panels assume $\mu=10^{4}$ km$^{-1}$ and $\lambda$ unchanged. Both numerical simulations clearly show an enhancement of flavor conversions in the three flavor case as compared to the two flavor case. In addition, the onset of  flavor conversion happens exactly at the same time in two and three flavors cases. 
However, the flavor instability grows faster for the larger value of $\mu$; it could be naively expected that this would lead to  more flavor conversion, if not the same, in the nonlinear regime. However, the opposite is true,  as shown in Fig.~\ref{fig:fig2}. 

 It is not clear from the time interval considered in our simulations in Fig.~\ref{fig:fig2}  whether a steady-state configuration is reached. In fact, as the time over which the simulation is run  increases, a larger number of angle bins is required, restricting the time over which the simulations can be performed. However, the main goal in this work is to gain insight on the transition to the non-linear phase in the two and three flavor regimes.

The reason for the  difference between two and three flavor evolution can be understood by looking at  the off-diagonal elements of the density matrices. The inset on the bottom of Fig.~\ref{fig:fig2}   shows the angle averaged evolution of the absolute values of $|\rho_{e\mu}|$ and $|\rho_{e\tau}|$. They  are almost identical in the linear regime where the growth of these quantities is exponential. 
 In the two flavor approximation, the growth rate of the absolute value of the off-diagonal element is the same as $|\rho_{e\mu}|$ and $|\rho_{e\tau}|$ in the linear regime, 
and $|\rho_{e\mu} - \rho_{e\tau}| = 0$. However, in three flavors, Fig.~\ref{fig:fig2} shows that  $|\rho_{e\mu} - \rho_{e\tau}|$ and $|\rho_{\mu\tau}|$ grow exponentially, breaking  the symmetry between $\nu_{\mu}$ and $\nu_{\tau}$ sectors. These findings highlight that the fast flavor evolution is intrinsically different in three flavors from the two flavor case. The stark difference between the two flavor and three flavor cases was not reported  in Ref.~\cite{Chakraborty:2019wxe} since it can be captured by the linear stability analysis only on a sub-leading level due to  the growth of $|\rho_{e\mu} - \rho_{e\tau}|$, while it becomes very evident in the non-linear regime.

Our findings highlight that  the fast flavor evolution of two and three flavor systems can be significantly different even in the case of trivial initial angular distributions for the non-electron type neutrinos, differently from what pointed out in Ref.~\cite{Capozzi:2020kge}. It should be noted that, although we have solved the system of equations including the matter term, 
our results are not sensitive to the exact value of the matter potential, as shown in Appendix~\ref{sec:matter}.

\subsection{Case I: Dependence of the flavor conversion physics on the vacuum mixing parameters}
\label{vacuum}
\begin{figure*}
\includegraphics[width=0.49\textwidth]{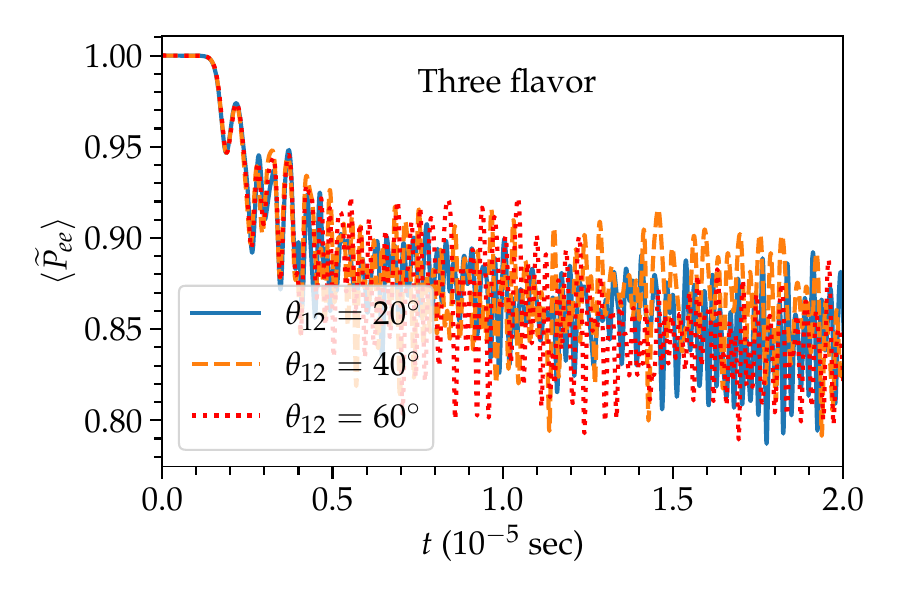}
\includegraphics[width=0.49\textwidth]{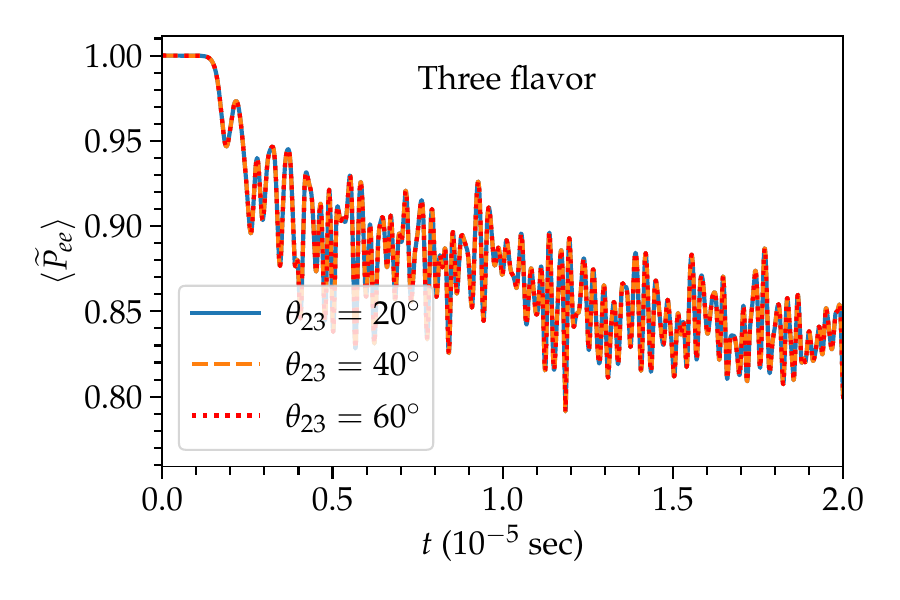}
\caption{Temporal evolution of $\langle \widetilde{P}_{ee}\rangle$ for Case I  for three different values of $\theta_{12}$ (left) and $\theta_{23}$ (right), respectively. A small quantitative dependence of $\langle \tilde{P}_{ee}\rangle$ on $\theta_{12}$ is visible, although the qualitative behavior is unchanged. Absolutely no dependence on $\theta_{23}$ is evident instead.}
\label{Figvac}
\end{figure*}

The dramatic difference between the neutrino conversion physics in  two and three flavors naturally raises  questions on the sensitivity of the three flavor evolution to the specific choice of the mixing parameters. To this purpose, in Fig.~\ref{Figvac},  we explore the dependence of $\langle \widetilde{P}_{ee}\rangle$ on  $\theta_{12}$ (left) and $\theta_{23}$ (right). The temporal evolution of the three flavor system is fairly independent on the vacuum mixing parameters. However, minor quantitative differences due to the exact value of $\theta_{12}$ are visible in the left panel of Fig.~\ref{Figvac}. Irrespective of the value of $\theta_{12}$,  rapid oscillations develop over time and the survival probability reaches smaller values than the one in the two flavor case shown in Fig.~\ref{fig:fig2}. The dependence of $\langle \widetilde{P}_{ee}\rangle$ on $\theta_{23}$ (right panel of Fig.~\ref{Figvac}) is negligible, as expected~\cite{Akhmedov:2004ny,Fogli:2008fj}. 

We have also verified that our results  are qualitatively insensitive to all  other vacuum mixing parameters, including the $\mathcal{CP}$-phase $\delta$ (results not shown here). 
The role of the vacuum mixing angles is to provide seeds for the non-linear growth. The qualitative features of the system are, however, independent on the seeds. In addition, these findings  confirm the high degree of numerical accuracy of our calculations.

\subsection{Case II: Crossings in the $\mu$ and $\tau$ sectors}
\begin{figure*}
\includegraphics[width=0.49\textwidth]{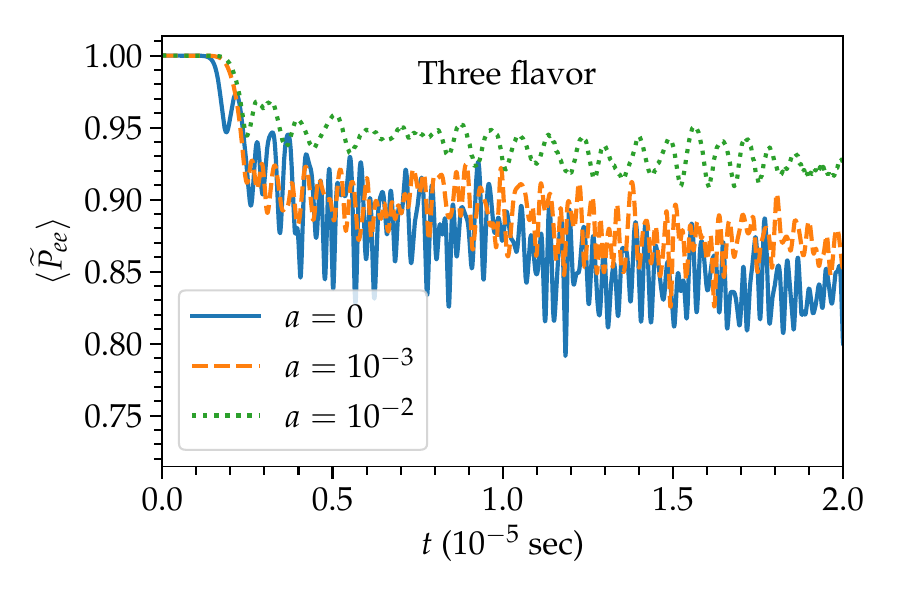}
\includegraphics[width=0.49\textwidth]{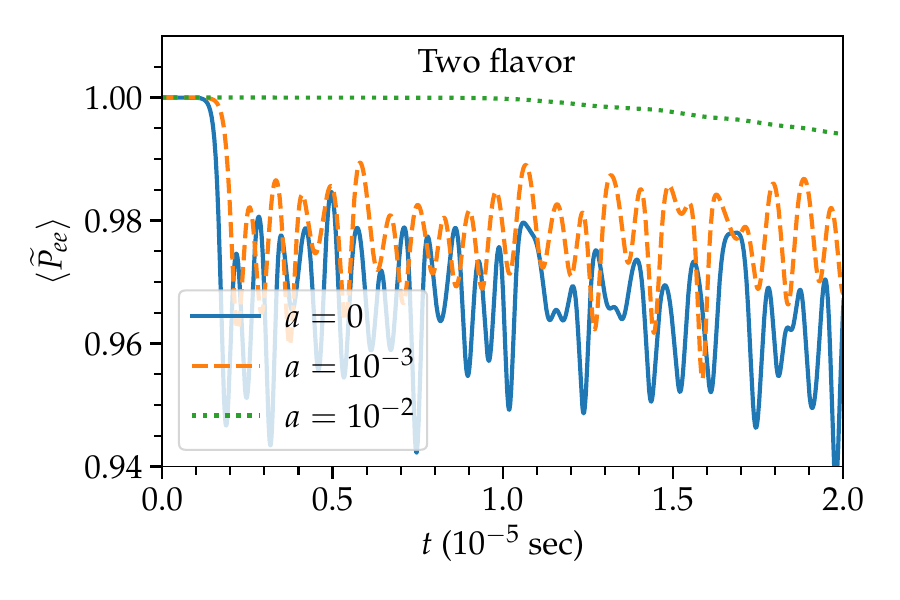}
\caption{{\it Left:} Temporal evolution of $\langle \widetilde{P}_{ee}\rangle$ for various values of the parameter $a$ (see Eq.~\ref{mucr2}). The high frequency modulations of $\langle \widetilde{P}_{ee}\rangle$  vanish as the magnitude of the nELN crossing exceeds a certain threshold (i.e., $a$ grows) and  the three-flavor enhancement of  conversions  disappears.  {\it Right:} The same as the left panel, but for the two flavor case. Also in the presence of nELN crossings, the two flavor case displays a different flavor outcome than the three flavor one.}
\label{Fig3}
\end{figure*}
As discussed in Refs.~\cite{Capozzi:2020syn,Capozzi:2020kge}, nELN crossings can occur in core-collapse supernovae and, in principle, can trigger additional flavor mixing. 
We  explore the impact of $\mu$ neutrino lepton number crossings  by setting up the system with initial conditions defined by Eqs.~\ref{cases1}, \ref{cases2}, \ref{mucr1}, and \ref{mucr2}.

Figure~\ref{Fig3} shows the angle averaged ratio between the initial and final content of $\nu_e$  for  $a = 0$, $10^{-3}$, and $10^{-2}$ in three (two) flavors on the left (right) panel. 
For small values of $a$, we do not see any substantial difference with respect to the $a=0$ case; however,  for $a=10^{-2}$, the symmetry between non-electron neutrinos and antineutrinos  is sufficiently broken. This leads to a disappearance of the enhancement of the flavor mixing caused by three flavor effects (see also Fig.~\ref{fig:fig2} for comparison). Also in the presence of nELN crossings, the two flavor case displays a different flavor outcome than the three flavor one. 
The scenario with large $a$ ($a=10^{-2}$) captures the kind of effect also discussed in Ref.~\cite{Capozzi:2020kge} due to different angular distributions for the non-electron type neutrinos, in spite of the differences being much smaller in magnitude compared to those considered in Ref.~\cite{Capozzi:2020kge}.

Our findings clearly show that  nELN crossings,  if  much smaller in magnitude than the ELN one, negligibly affect the three  flavor evolution. Due to the overwhelming excess of electrons over muons in supernovae as well as neutron star mergers, the difference between $\nu_{\mu}$ and $\bar{\nu}_{\mu}$ is bound to be far less substantial than the one between $\nu_{e}$ and $\bar{\nu}_{e}$~\cite{Bollig:2017lki,Capozzi:2020syn}; hence, one should realistically expect scenarios similar to the ones discussed for small values of $a$.

It should be noted that the two flavor calculation in Ref.~\cite{Capozzi:2020kge} is performed by adopting  an initial ensemble of $\nu_e$ and $\bar\nu_e$ only, while the three flavor calculation includes   non-trivial initial angular distributions for the non-electron flavors~\cite{pricom}. This approach is not consistent since it is equivalent to consider the evolution of two systems with a different number of particles at $t=0$. In addition, it effectively changes the self-interaction potential strength and obfuscates the effect caused by the three flavor evolution due to the inherent differences determined by the non-trivial initial angular distributions of the non-electron type neutrinos. In the light of these differences, a direct comparison of our findings with the ones of Ref.~\cite{Capozzi:2020kge} is not possible.

\section{The effect of collisions}
\label{sec:collisions}
The flavor evolution in the presence of matter is not only affected by the matter effect, but also by collisions~\cite{Shalgar:2020wcx,Martin:2021xyl,Capozzi:2018clo}. In particular,  contrary to conventional wisdom,  collisions can lead to an enhancement of flavor conversions in certain circumstances~\cite{Shalgar:2020wcx}. Following Ref.~\cite{Shalgar:2020wcx}, we   investigate whether collisions lead  to an enhancement of  flavor conversion in the three flavor framework.

We follow the same formalism as the one proposed in Ref.~\cite{Shalgar:2020wcx} and modify Eq.~\ref{eq:eom1}   to include direction changing collisions:
\begin{eqnarray}
\frac{d \rho(\cos\theta)}{dt} = &-& \int_{-1}^{1} \Closs \rho(\cos\theta) d\cos\theta^{\prime}\nonumber\\
&+& \int_{-1}^{1} \Cgain \rho(\cos\theta^{\prime}) d\cos\theta^{\prime}\nonumber\\
&-& i[H(\cos\theta),\rho(\cos\theta)]\ ,\nonumber\\
\frac{d \bar{\rho}(\cos\theta)}{dt} = &-& \int_{-1}^{1}\Clossbar \bar{\rho}(\cos\theta) d\cos\theta^{\prime}\nonumber\\
&+& \int_{-1}^{1} \Cgainbar \bar{\rho}(\cos\theta^{\prime}) d\cos\theta^{\prime}\nonumber\\
&-& i[\bar{H}(\cos\theta),\rho(\cos\theta)]\ ,
\label{eoms}
\end{eqnarray}
where we assume that that $\Closs = \Cgain = \Clossbar = \Cgainbar$, implying particle number conservation as well as the same cross-section for neutrinos and antineutrinos. The inverse of the mean free path of the neutrinos can be expressed in terms of the collision terms by $\mathcal{C} = 2 \Closs = 2 \Cgain = 1$~km$^{-1}$ (see Appendix~\ref{sec:convergenge} for details on the numerical convergence).

\begin{figure}
\includegraphics[width=0.49\textwidth]{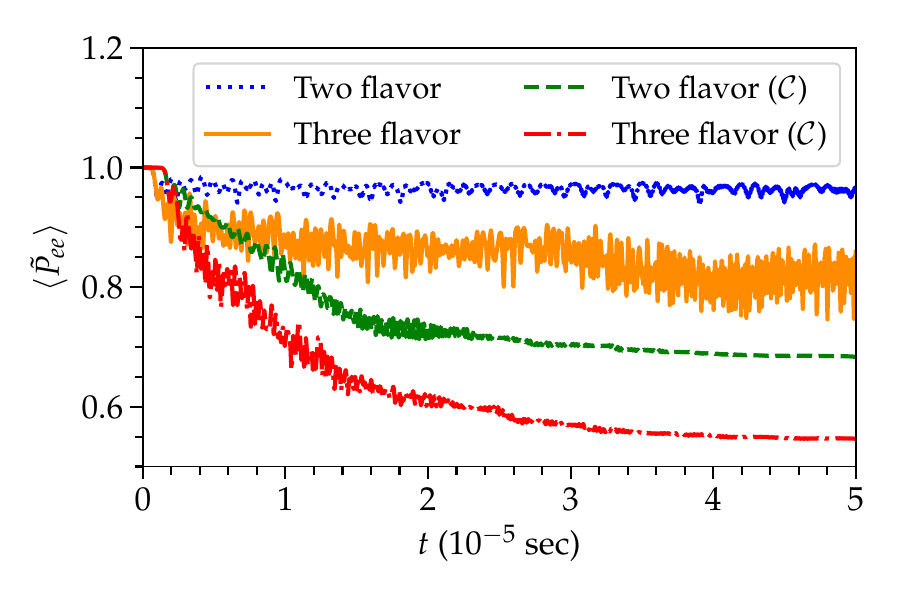}
\caption{Temporal evolution of of $\langle \widetilde{P}_{ee}\rangle$  for the two and three flavor scenarios, with and without collisions. 
 Collisions have the effect of further enhancing flavor mixing  in two and three flavors.}
\label{fig:collisions}
\end{figure}

As displayed in Fig.~\ref{fig:collisions}, obtained by relying on the initial conditions of Case I, the inclusion of collisions leads to an additional  enhancement of flavor conversions on top of the one due to three flavor effects. This confirms the findings of Ref.~\cite{Shalgar:2020wcx} and it is due to the fact that collisions redistribute the flavor content across angular bins creating more favorable conditions for flavor mixing. 

\section{Conclusions and outlook}
\label{sec:conclusions}
The consequences on flavor mixing of fast pairwise flavor conversion remain to be understood. 
In this paper, we compare the flavor outcome within the usually adopted two flavor approximation to the one resulting from the full three flavor case. 

The two flavor approximation  leads to differences with respect to the three flavor case  already in the linear regime and such differences become amplified  in the non-linear regime. The  exponential growth of  $|\rho_{e\mu} - \rho_{e\tau}|$ in three flavors (which  remains zero in the two flavor approximation  by definition)  leads to a symmetry breaking  between the $\nu_{\mu}$ and $\nu_{\tau}$ sectors that strongly affects the flavor conversion physics in the non-linear regime. This is a crucial difference at the essence of fast flavor mixing which was not pointed out in Refs.~\cite{Capozzi:2020kge}.

The most vital component that leads to the failure of the two flavor approximation in fast flavor conversions is the nearly identical growth rate of $|\rho_{e\mu}|$ and $|\rho_{e\tau}|$; the latter is, however, seeded by a slightly different initial perturbation due to the vacuum Hamiltonian in three flavors.

Recent work has pointed out that the inclusion of muons in hydrodynamical simulations can lead to differences between the distributions of the non-electron flavors and induce crossings in the muon neutrino lepton number distributions~\cite{Bollig:2017lki,Capozzi:2020syn}. Although our findings are in qualitative agreement with the ones of Ref.~\cite{Capozzi:2020kge},  this work shows that crossings in the angular distributions of non-electron type neutrinos are not crucial to appreciate an enhancement of flavor mixing in the full three flavor scenario. Notably, we find more extreme effects for smaller crossings in the non-electron flavors~\cite{Bollig:2017lki,Capozzi:2020syn} on the flavor conversion probability, which have not been considered in Ref.~\cite{Capozzi:2020kge}.
Finally, the enhancement of flavor mixing in three flavors is further exacerbated by collisions.

In conclusions, it is  vital to our understanding of the  neutrino flavor evolution  in any astrophysical system to include all three neutrino flavors.  Our findings  reinforce the limitations of the linear stability analysis as  the conditions leading to the growth of flavor instabilities may  seem identical for some components of the density matrix in two and three flavor systems;  however the flavor conversion physics  is inherently different in the non-linear regime. 

\acknowledgments
We would like to thank Madhurima Chakraborty and Sovan Chakraborty for useful discussions  and comments on the manuscript. We are grateful to the Villum Foundation (Project No.~13164), the Danmarks Frie Forskningsfonds (Project No.~8049-00038B),  and the Deutsche Forschungsgemeinschaft through Sonderforschungbereich
SFB~1258 ``Neutrinos and Dark Matter in Astro- and
Particle Physics'' (NDM).


\appendix 
\section{Numerical convergence of the results}
\label{sec:convergenge}
The numerical results presented in  this paper were obtained by carrying out simulations obtained by discretizing over the angular bins and numerically evolving Eqs.~\ref{eq:eom1}  for each angular bin. In order to do so, we rely on the Boost  implementation of the eighth order Runge-Kutta-Fehlberg method with an adaptive step-size~\cite{BoostLibrary}. To avoid any numerical error, we perform the numerical simulations with extremely tight absolute and relative tolerances of $10^{-16}$ and $10^{-12}$, respectively. 

\begin{figure}[]
\includegraphics[width=0.49\textwidth]{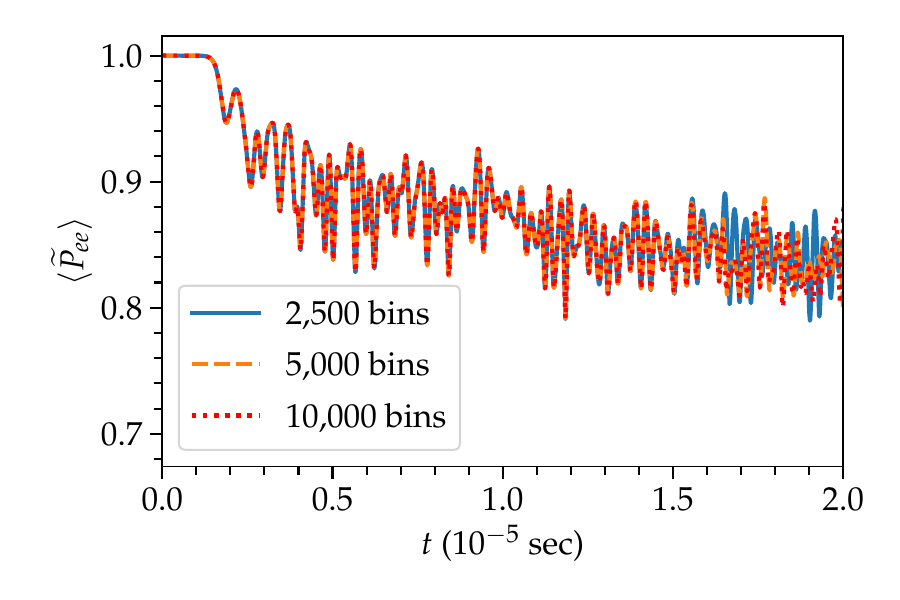}
\caption{Temporal evolution of $\langle \widetilde{P}_{ee}\rangle$ for Case I for 2,500, 5,000, and 10,000 angle bins. $\langle \widetilde{P}_{ee}\rangle$ does not depend on the number of angular bins until $t=1.5 \times 10^{-5}$~s, after that the results for 2,500 and 5,000 angular bins slightly diverge from each other. However, a comparison between the results obtained by employing  5,000 and 10,000 angular bins shows that 5,000 angle bins are sufficient to achieve convergence. 
}
\label{Figconv}
\end{figure}

In addition, we have ensured that the number of angular bins is sufficient to reach convergence by solving Eqs.~\ref{eq:eom1} with a different number of angular bins. 
For all the configurations considered in this paper, 5,000 angle bins have been tested to be sufficient to achieve convergence over the time scales of interest. An example is reported in Fig.~\ref{Figconv}, where the temporal evolution of $\langle \widetilde{P}_{ee}\rangle$  for Case I is shown for three different for  2,500, 5,000, and 10,000 angular bins.

It should be noted that the number of angle bins required to achieve numerical convergence depends on the timescale over which the equations of motion are evolved. As $t$ increases, the angular distributions acquire structures on  smaller scale. Consequently, for a given system configuration, the number of angular bins required for convergence increases as the time over which the system is evolved increases. This can be clearly seen in Fig.~\ref{Figconv}. The numerical results for 2,500 and 5,000 angle bins are identical up to  $\simeq 1.5 \times 10^{-5}$~s, after which they start deviating from each other. However, the employment of 5,000 angle bins guarantees convergence; in fact the evolution of $\langle \widetilde{P}_{ee}\rangle$ is identical to the one obtained by using 10,000 angular bins. 

The inclusion of collisions smoothes out some of the angular structure and the number of angle bins required for a system with collisions is smaller than the one without collisions, as discussed in Ref.~\cite{Shalgar:2020wcx}; in our case, 1,000 angular bins were enough to guarantee numerically convergent results (related plots not shown here).

\section{Matter effects}
\label{sec:matter}
In the two flavor approximation, the matter term has the effect of reducing the effective mixing angle while keeping the effective vacuum frequency unchanged~\cite{EstebanPretel:2008ni}. It is thus possible to take into account the effect of the matter term by using a small effective mixing angle and keeping the vacuum frequency unchanged. However, it is not obvious whether this strategy can work in the three flavor approximation. To this purpose we set the initial conditions for the neutrino ensemble as in Case I and consider two different values of the matter potential, $\lambda = \mu$ and $\mu/100$. 

\begin{figure}[b]
\includegraphics[width=0.49\textwidth]{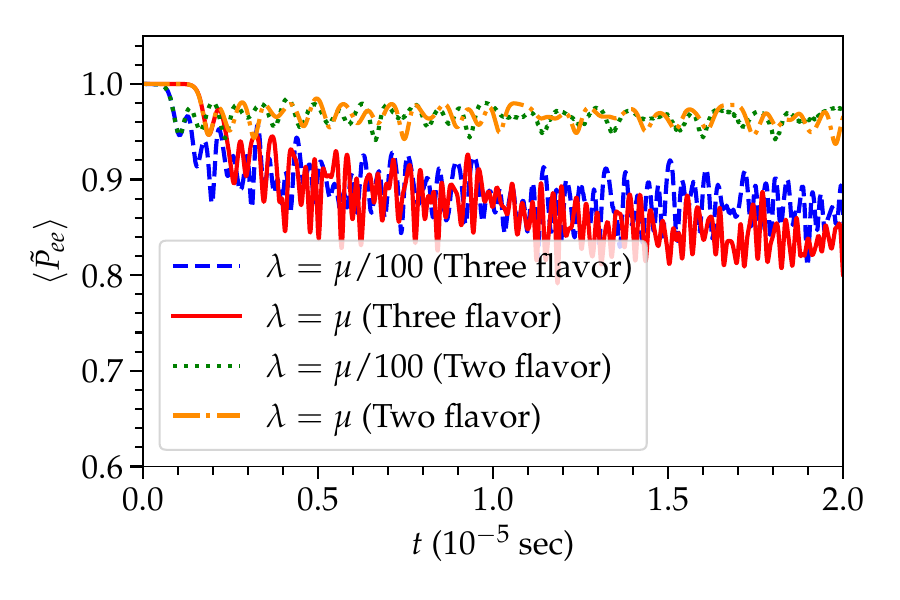}
\caption{Temporal evolution of $\langle \widetilde{P}_{ee}\rangle$  in two and three flavors  with $\mu=10^{5}$ km{-1} and $\lambda = \mu$ and $\mu/100$. The blue dashed line shows the temporal evolution in the case of a matter potential that is smaller than the one expected in astrophysical scenarios, while the solid red line shows the flavor evolution for $\lambda = \mu$; the kind of matter potential expected in a realistic astrophysical system. The exact value of $\lambda$ does not affect the final flavor outcome drastically, but affects the onset of flavor conversions.}
\label{Fig5}
\end{figure}

Figure~\ref{Fig5} shows the temporal evolution of $\langle \widetilde{P}_{ee}\rangle$  in two and three flavors for different values of $\lambda$. We can see that, although the features on a small time scale are affected by the matter term, the trend is unaffected by the matter term on a larger time-scale. 

 \bibliography{threef.bib}
\end{document}